\documentclass{hcj}

\journalyear{2015}
\journalvolume{1}
\journalissue{1}
\journalpages{101-131}
\articledoi{10.15346/hc.vxix.x}
\journalcopyright{\copyright{} 2015, Hogg \& Lerman. CC-BY-3.0}
\usepackage{graphicx}
\usepackage{amssymb,amsfonts}%,amsmath} %amsmath gives warning, apparently due to use of \choose

\newcommand{\eq}[1]{Eq.~(\ref{eq.#1})} % Ref. to equation
\newcommand{\fig}[1]{Fig.~\ref{fig.#1}}

\newcommand{\tbl}[1]{Table~\ref{table.#1}}

\newcommand{\sect}[1]{Section~\ref{sect.#1}}
\newcommand{\sectA}[1]{Appendix~\ref{sect.#1}}
\newcommand{\sectlabel}[1]{\label{sect.#1}}
\newcommand{\eqlabel}[1]{\label{eq.#1}}
\newcommand{\figlabel}[1]{\label{fig.#1}}
\newcommand{\tbllabel}[1]{\label{table.#1}}
\newcommand{\view}{\mbox{view}}
\newcommand{\vote}{\mbox{vote}}
\newcommand{\click}{\mbox{click}} % url click
\newcommand{\Lfull}{L_{\mbox{\scriptsize full}}}
\newcommand{\model}[1]{{\sc #1}}

\newcommand{\figwidth}{3.0in}
\newcommand{\halffigwidth}{2in}

\newcommand{\remove}[1]{}
\usepackage{color}

\title{Disentangling the Effects of Social Signals}
\author{
Tad Hogg\affil{Institute for Molecular Manufacturing}
        \and
Kristina Lerman\affil{USC Information Sciences Institute}
        }
\authorrunning{T.~Hogg and K.~Lerman}

\begin{document}

\maketitle
\begin{abstract}
Peer recommendation is a crowdsourcing task that leverages the opinions of many to identify interesting content online, such as news, images, or videos. Peer recommendation applications often use social signals, e.g., the number of prior recommendations, to guide people to the more interesting content. How people react to social signals, in combination with content quality and its presentation order, determines the outcomes of peer recommendation, i.e., item popularity. Using Amazon Mechanical Turk, we experimentally measure the effects of social  signals in peer recommendation. Specifically, after controlling for variation due to item content  and its position, we find that social  signals affect item popularity about half as much as position and content do. These effects are somewhat correlated, so social  signals exacerbate the ``rich get richer'' phenomenon, which results in a wider variance of popularity. Further, social signals change individual preferences, creating a ``herding'' effect that biases people's judgments about the content. Despite this, we find that social  signals improve the efficiency of peer recommendation by reducing the effort devoted to evaluating content while maintaining recommendation quality.
\end{abstract}

\section{Introduction}
Every day people make a multitude of decisions about what to buy, what to read, where to eat, and what to watch. These decisions are often influenced by the actions of others. The interplay between individual choices and collective action is responsible for much of the  complexity of markets~\cite{Salganik06} and social behaviors~\cite{Ratkiewicz10}. This is especially evident in peer production web sites, such as YouTube, Reddit, and Tumblr.
To help users identify interesting items, these web sites often highlight a small fraction of their content based on the reactions of prior users. Such reactions include whether the user clicked on an item, downloaded it, or recommended it; actions that provide implicit opinions of content quality. Some web sites also allow users to explicitly rate content.
The web sites may then showcase highly-rated or popular content on a special web page or rank order items according to these ratings. In addition, the web sites can use the ratings as a \emph{social signal} by showing the users these ratings, e.g., the number of votes, likes or recommendations prior users gave to the content. This type of social signal reflects the opinions of the user community as a whole, and is distinct from signals a user might receive from friends.
%This paper
%focuses on the former type of social signal, examining
%examines how this information about general community preferences influences user behavior.
Understanding how social signals conveying information about community preferences influence user behavior could enable web designers to predict behavior, optimize it, and steer it towards desired goals.

The implicit (e.g., through ranking) and explicit (through social signals) information about others' preferences together affect people's decision-making processes, an interaction not resolved by previous experimental studies~\cite{Salganik06,Lorenz11,Muchnik13}. Moreover, social signals can have multiple effects.
By conveying information about which items others found interesting or liked, such signals could simply direct a person's attention to items he or she may also find interesting~\cite{krumme12}. Alternately, as with peer pressure and social norms, social signals could change individual's preferences, leading him or her to find some items more interesting than in the absence of social signals.
In practice, these and other effects, such as novelty and content quality, combine to affect what content is recommended.

We disentangled some of these effects through experiments on peer recommendation conducted using Amazon Mechanical Turk. We showed people a list of science stories and asked them to vote for, or recommend, ones they found interesting. We measured the degree to which displaying the number of prior votes, which acts as the {social signal}, affects the votes stories receive, after controlling for their content and position within the list shown to users. We also studied how social signals affect individual behavior, specifically, which stories people choose to evaluate and their preferences for them.

% summary of rest of paper
After describing related work, we present our experimental design and a model used to estimate user behaviors we cannot measure directly. We then present the experimental results and summarize their implications for designing peer recommendation systems. The appendix provides details of the experiment setup, user responses to the social signals and evaluation of the model.

\section{Related Work}

Social influence helps a group to achieve consensus and adopt new ideas~\cite{KatzLazarsfeld,Rogers03,Bond12}. However, it introduces a bias into judgements that may skew collective outcomes~\cite{Salganik06,Lorenz11,Muchnik13}. This bias creates an ``irrational herding'' effect~\cite{Muchnik13} that amplifies small differences in individual response, leading to a ``rich get richer'' phenomenon. Experimental studies of cultural markets concluded that social influence signals  lead to large inequality and unpredictability of collective outcomes~\cite{Salganik06,Salganik09}.
However, before information about others' preferences, can influence a person, he or she must first see it. Due to human cognitive biases, the presentation order of items strongly affects how people allocate attention to them~\cite{Payne51,Buscher09}. In an earlier experimental study~\cite{lerman14as}, we showed that ordering alone could explain much of the unpredictability and inequality of outcomes that is attributed to social influence. In this study, we quantify the effects of social influence through web-based experiments.

Several previous studies attempted to quantify the effects of influence of social signals by modeling online behavior data. In particular, some studies use observational data from well-established social media sites to estimate parameters of stochastic models that include various types of influence on how users react to the presentation of content. For instance, studies of news aggregators~\cite{Hogg12epj,Stoddard15} and product recommendation~\cite{Wang14} used such models to identify contributions of position bias, social signals and the appeal of the content to popularity.
%A different example of social influence is a study including the social network relationship between a user and the content's creator on the site~\cite{Hogg12epj}. Instead of reacting to the number of prior votes, this study indicates users react differently depending on \emph{who} votes.
These studies not only indicate relative importance of social signals, but are also useful in predicting the growth in popularity for new content based on the early responses to that content.

These observational studies can identify correlations but do not directly test self-selection effects, causal influences and counterfactual outcomes. Nevertheless, observational studies offer significant advantages of large scale and users performing real-world tasks on social media sites. Thus these studies complement randomized experiments, such as MusicLab~\cite{Salganik06} and the experiments presented in this paper, which are smaller and present tasks in a more stylized setting.

Identifying the extent to which the simplified scenarios of experiments generalize to actual web behavior is an important issue for evaluating experiments. When a sufficient number of observational and experimental studies are available on a particular question, comparison can directly test generalization, e.g., quantifying the effect of peer influence on worker productivity~\cite{herbst15}. A similar approach could apply to evaluating studies of social influence in crowdsourcing once enough studies are available.

\section{Experiment Design}

\begin{figure}[t]
\centering  \includegraphics[width=\figwidth]{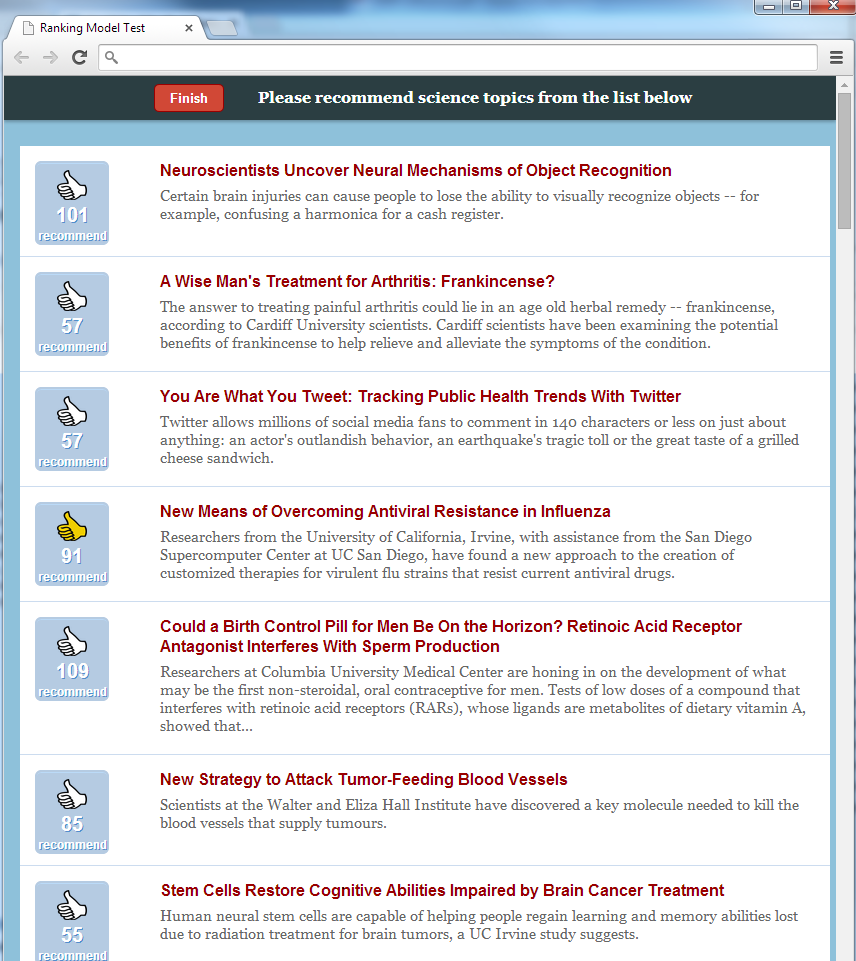}
\caption{Screenshot of a web page shown during an experiment. The participant clicks on the button to the left of a story's summary to recommend that story. The buttons include the number of prior participants who recommended each story. The colored graphic next to the fourth story indicates the participant has recommended, or \emph{voted for}, that story.
% UI and this caption uses 'recommend' whereas text of paper uses 'vote'
}
\figlabel{screenshot}
\end{figure}

We studied peer recommendation with randomized experiments on Amazon Mechanical Turk (Mturk), which is a popular platform for experimental behavioral research~\cite{bohannon11,mason12,crump13,Kittur13}. Our experiments presented a list of science stories to people and asked them to vote for stories they found interesting. This voting is similar to ``liking'' or recommending items in social media web sites.

%\noteTH{added motivation to use science stories, instead of a mix of topics (questioned by WSDM reviewer)}
We used science stories to have a single general topic area, thereby reducing variation that could arise from a mix of topic areas. In effect, the experiment mimics the behavior of a special interest subgroup of a web site presenting a wide variety of content types. Moreover, science articles are less time-sensitive than, say, current news or sports stories. This allowed us to run the experiment with different users over several months without measurable change in user response to the stories.

The interface, shown in \fig{screenshot}, displayed the title and summary of each story. The title was linked to the full story via its url. Thus people could read the full story by clicking on the title, but were not required to do so. We refer to such actions as ``url clicks''.
% IRB mentioned in supplementary information
For the influence experiments, the interface displayed the number of prior votes each story received, starting from zero for all stories. This number provided the \emph{social signal}. The user recruitment and vetting procedures are described in \sectA{vetting}.

Prior experiments~\cite{lerman14as} measured how story position affects how many votes it receives and compared several policies for ordering the stories. For the present study with the social signals, we used two of these ordering policies:
% -- combine discussion of all orderings into one paragraph to save space
% WSDM reviewer 2 suggested separating the discussion of orderings
\begin{itemize}
\item the \emph{fixed} ordering showed all stories in the same order to everyone. This corresponds to the common practice on web sites of showing stories in the same, e.g., chronological, order to all users to emphasize recent additions.

% We use single word 'activity' for this ordering throughout the paper
\item the recency of \emph{activity} ordering presented stories in chronological order of the latest vote they received, with the story with the most recently vote at the top of the list. This is similar to Twitter's policy of displaying the most recently tweeted or retweeted post at the top of the followers' streams.
%By continually moving the most recently voted for stories to the top of the list, the activity policy creates a dynamic ordering that allows us to contrast competing effects of position bias and social influence.
\end{itemize}
In addition, as a control condition without the social signal, the \emph{random} ordering presented the stories in a new random order for each user, without showing the number of prior votes.

The social signal shown to a given user depends on the actions of prior users. Thus repeating the experiment with same ordering policy with different users can produce different signals, which may also affect story order in the activity-based ordering. To evaluate the significance of this variation, we performed two independent, or ``parallel worlds'', experiments for each ordering policy. Each parallel world starts from the same initial state: each story starting with zero votes and shown in the same order as the fixed ordering.

\section{Effect of Content and Position}

Stories vary in how interesting they are to users, and users vary in their effort to evaluate content. Most users in our experiments chose stories based solely on the title and summary.
To quantify this behavior, we operationally define the \emph{appeal} $r_s$ of a story $s$ as the conditional probability a user votes for the story $s$ given the user has viewed its title and summary: $\Pr(\vote|\view)=r_s$.

Our experiments do not indicate which stories users view, so we cannot directly measure the probability a user views a story, $\Pr(\view)$, and hence we do not directly estimate $\Pr(\vote|\view)$. Instead, we use a model to jointly estimate views and story appeal.

\subsection{Stochastic Model of Voting}
\sectlabel{model}
To separate the effects of ordering and story content from that of social signals, we developed a model of user behavior in the absence of social influence.
%This section describes the model, and \sectA{evaluations} discusses measures of how well it describes behavior without the influence signal.
To vote for a story, a user has to both see it and find it appealing. Our model accounts for these two factors: the probability to view the story at position $p$, $\Pr(\view)=v_p$, and its appeal $\Pr(\vote|\view)=r_s$. The model jointly estimates these factors from the observed votes.

The model makes several simplifying assumptions. Specifically, we assume a homogeneous user population, with no systematic differences between users in their preference for stories or how they navigate the list of stories. Therefore, the probability to view a story depends only on its position $p$ in the list. Moreover, the probability a user votes for a story after viewing it depends only on the story $s$ but not the user or the story's position in the list.
While there may be variations among user preferences, e.g., for technology or medicine, we focus on average behavior of users as the primary effect for peer recommendation in our experimental context. The model does not include social signals and their effects on users.

In addition, we assume that viewing each story is an independent choice by the user. This assumption was also used in a model of the Salganik et al.~experiments~\cite{krumme12}. This contrasts with models of list navigation (see e.g.,~\cite{Huberman:1998eq,Craswell08}) which posit that users view stories in order in a list until they decide to quit, so that viewing a story at position $p$ means all stories at prior positions were also viewed. While we do not observe such sequential navigation in user actions, there is significant dependence in acting on stories near a previously action, which our model does not capture.
Another independence assumption is that users consider each viewed story independently. This contrasts, for example, to a dependence arising from users quitting after voting for a set number of stories, in which case, whether a user views a story depends on the number of votes that user has already made.

Due to these assumptions, our model does not capture detailed behavior of individual users. Nevertheless, as shown in \sectA{evaluations}, the model learned from one experiment correctly predicts the behavior of users in new experiments.
Thus our simplifying assumptions allow understanding the aggregate effects of story ordering.

Specifically, when story $s$ is presented to a user at position $p$ in the list (where $p$ ranges from 0 to $S-1$, where $S=100$ is the number of stories), we model the probability $\rho(s,p)$ that a user votes for the story as: % the product of its appeal $r_s$ and probability to view it at position $p$:
\begin{equation}\eqlabel{prob(recommend)}
\rho(s,p)=r_s v_p
\end{equation}
where
$r_s$ is the story appeal and $v_p$ its visibility at position $p$. As a reminder, appeal is defined as the conditional probability the user votes for the story given he or she views it: $r_s = P(\mbox{vote for s} | \mbox{view s})$. Similarly, visibility is defined as the conditional probability the user views the story given that it is presented to him or her at position $p$: $v_p = P(\mbox{view s} | \mbox{s presented at position p})$.
%is the \begin{eqnarray}
%v_p &=& P(\mbox{view $s$} | \mbox{$s$ presented at position $p$}) \\
%r_s &=& P(\mbox{vote for $s$} | \mbox{view $s$})
%\end{eqnarray}
We assume that $v_p$ does not depend on the particular story $s$.

\subsection{Parameter Estimation}

We did not directly observe values for $v_p$ and $r_s$ in the experiments. Instead we estimated them by maximum likelihood, a statistical method for identifying values for variables that best explain the data. We used data from the random ordering experiments to determine model parameters. %, since it averages over the effect of position. [averaging not needed for estimation since we record story positions for each user; instead random ordering is helpful by distributing stories so they all have some attention, from being in higher profile positions, to observe voting behavior for estimating appeal]
Specifically, with this model, the log-likelihood for user $u$ to vote for a set $S_u$ of stories is
\begin{equation}\eqlabel{log-likelihood}
L_u = \sum_{s \in S_u} \log(r_{s} v_{p_{s,u}}) + \sum_{s \notin S_u} \log(1 - r_{s} v_{p_{s,u}})
\end{equation}
with story $s$ shown to the user at position $p_{s,u}$, where it has visibility $v_{p_{s,u}}$.
For a set of users $U$, the log-likelihood for all their votes is $\Lfull = \sum_{u \in U} L_u$\remove{ with $L_u$ for user $u$ given by \eq{log-likelihood}}.

The values $r_s$ and $v_p$ enter the likelihood above as a product. Therefore, their values can be rescaled by an overall factor $\alpha$ without changing the likelihood: i.e., replacing $r_s \rightarrow \alpha r_s$ and $v_p \rightarrow v_p /\alpha$ for all stories and positions does not change the likelihood. To constrain the parameters, we chose a value of $\alpha$ which gave $v_0 =1$.

Numerically maximizing $\Lfull$ gives estimates for the model parameters, namely the appeal $r_s$ for each story and visibility $v_p$ for each position.
Moreover, expanding $\Lfull$ around its maximum provides estimated confidence intervals for these parameters.

% use of random ordering already mentioned at start of Model section, no need to repeat here
\remove{We used the random ordering policy with no social influence as the control condition training set for this estimation.}
To reduce fitting to noise, we used regularizers based on prior expectations that visibility changes smoothly with position in a list and the stories had similar appeal. We used 10-fold cross-validation to determine the amount of regularization~\cite{abu-mostafa12}. %This regularization smoothed the visibility values, had only a small effect on the appeal estimates and reduced the 95\% confidence intervals of the estimates by about 20\%.

\begin{figure}[t]
\centering
\begin{tabular}{@{}c@{}c@{}}
\includegraphics[width=\halffigwidth]{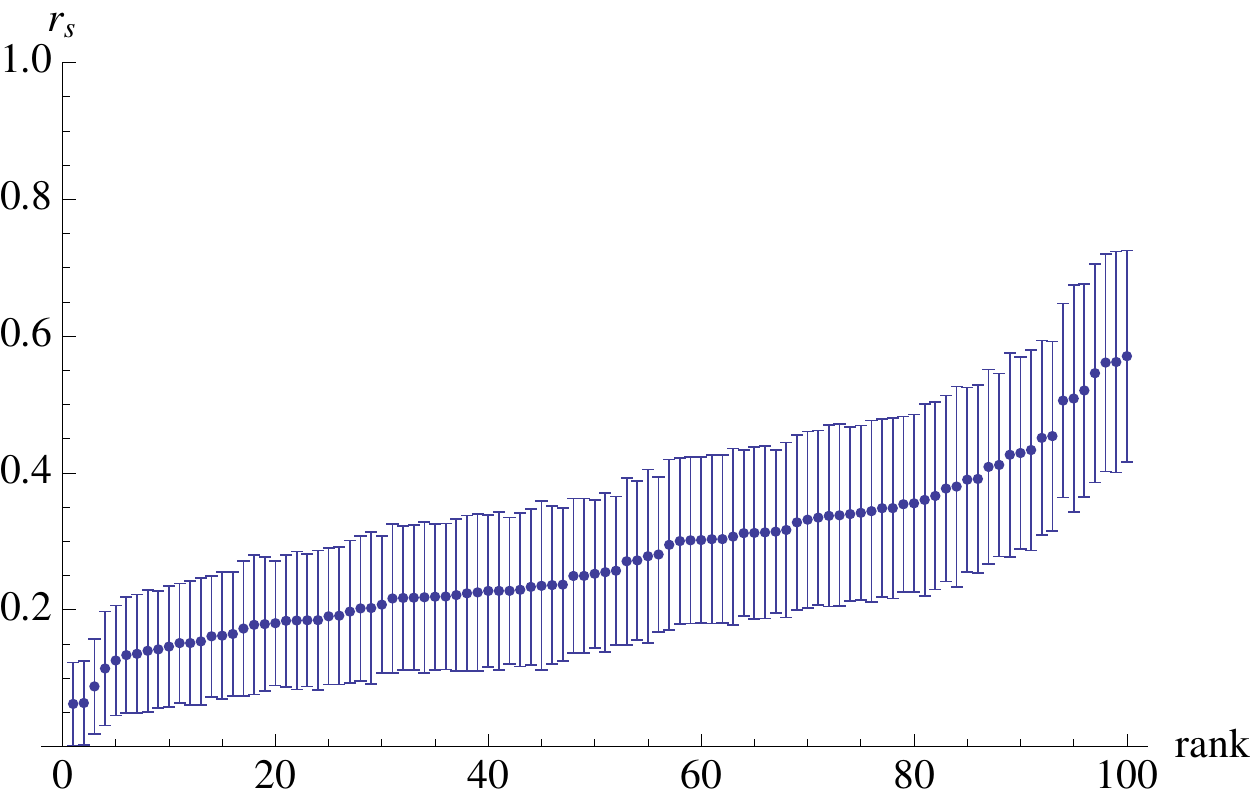}
&
\includegraphics[width=\halffigwidth]{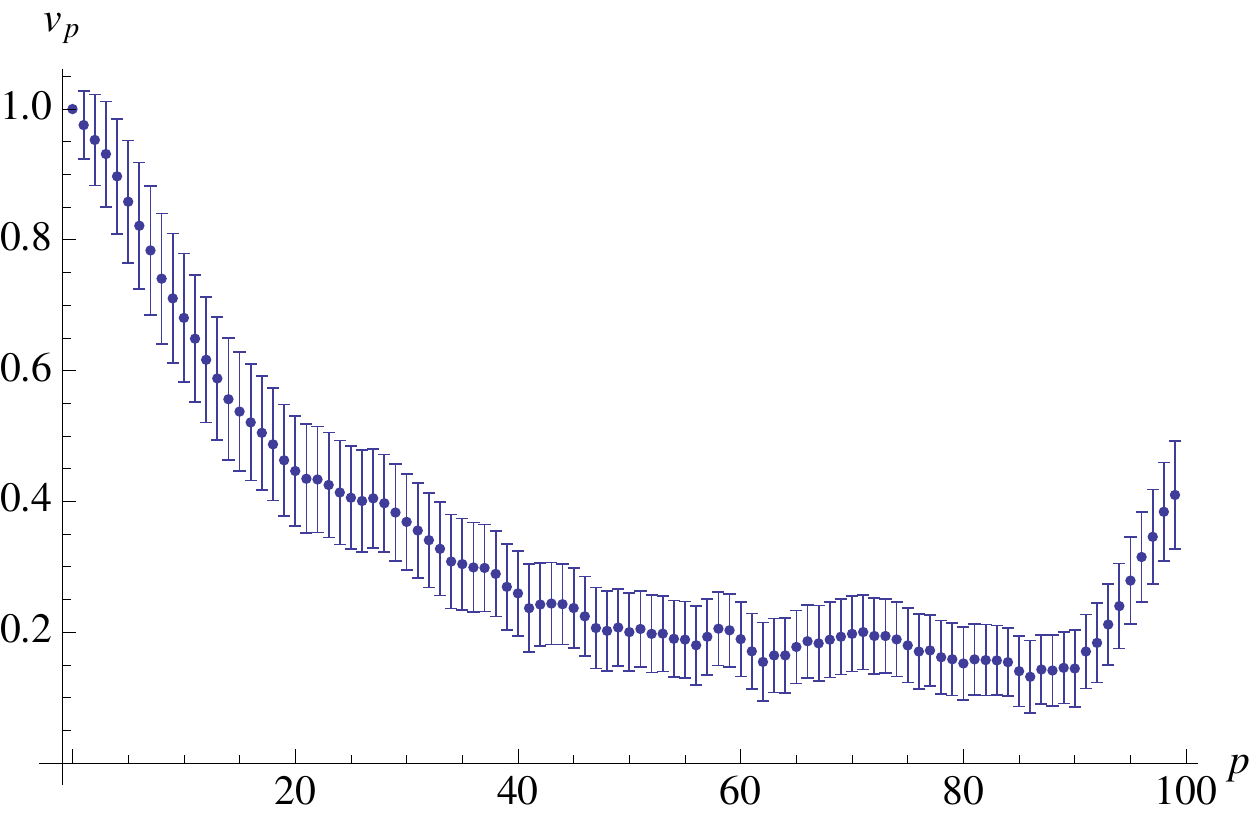}
\\
(a) & (b)
\end{tabular}
%\caption{Model estimates of story appeal, sorted in increasing order. Error bars give the 95\% confidence intervals.
\caption{Model estimates of story (a) appeal (sorted by rank) and (b) visibility (as a function of position $p$). Error bars give the 95\% confidence intervals. The value $v_0$ in (b) is set to 1 for overall scaling, so it has zero confidence interval.
}
\figlabel{interestingness}
\end{figure}

%\begin{figure}[!ht]
%\centering
%\includegraphics[width=\figwidth]{fig10}%eps-figs/visibility}
%\caption{Model estimates of story visibility\remove{, i.e., probability to view the story,} as a function of story position $p$. Error bars give the 95\% confidence intervals. The value $v_0$ is set to 1 by our choice of overall scaling, so it has zero confidence interval.}
%% alternative: instead of zero confidence interval for constrained value $v_0=1$, could evaluate deviations from max likelihood only in directions that maintain the scaling constraint, thereby getting intervals for all the v_p values (cf. "max likelihood with invariance.nb")
%\figlabel{visibility}
%\end{figure}

\fig{interestingness}(a) shows the $r_s$ estimates, indicating a five-fold variation in story appeal. This confirms that our experiment design qualitatively corresponds to the large variation in preferences for content  seen on social media web sites~\cite{Hogg12epj} and other domains, e.g., scientific papers~\cite{Barabasi13}.
\fig{interestingness}(b) shows the estimates for the probability to view the story $v_p$, which quantifies the position bias~\cite{Payne51} in our experiments. This quantity decreases rapidly with position: a story at the top of a list gets about five times as much attention as a story in the middle of the list, consistent with direct measurements reported in \cite{lerman14as}. A few users focus on the end of the list, resulting in the increase in $v_p$ for the last few positions in the list.

%\subsection{Estimating $\Pr(\click|\view)$}\sectlabel{P(click|view)}
%% using model to estimate $\Pr(\click|\view)$  -- used in discussion of "Effect of Content"
%
%We use the model to estimate $\Pr(\click|\view)$ for each story, i.e., the probability a user clicks on a story's url to read the full text of the story after viewing its title and summary.
%A user clicking on a story necessarily first views it, so $\Pr(\click|\view)=\Pr(\click)/\Pr(\view)$. We use the model's estimate of\remove{ the probability a user views a story at position $p$,} $\Pr(\view)=v_p$.  We define the indicator function $\chi(\click;u,s)$ to be 1 or 0 according to whether user $u$ clicks on story $s$. We use data from all our experiments without social influence~\cite{lerman14as} to estimate $\Pr(\click|\view)$ for story $s$ as the average over users of $\chi(\click;u,s)/v_{p_{s,u}}$ where $p_{s,u}$ is the position of story $s$ for user $u$.
%%\noteTH{This is a precise but wordy explanation of how we estimate $\Pr(\click|\view)$. An alternative would be to expand the model to jointly estimate three quantities: the two we already include $\Pr(\vote|\view)=r_s$, $\Pr(\view)=v_p$ and an additional $\Pr(\click|\view)$. But this would (presumably) only use the random interface experiment for estimation and hence have fewer data points than using all experiments to estimate $\Pr(\click|\view)$ with the method described in this paragraph}

\subsection{Url Clicks}
Some users put more effort into evaluating a story by clicking on its url to see the full story. In our experiments about 25\% of the users clicked on at least one url.
Voting for a story in this case involves two behaviors: the url click and the vote. The corresponding conditional probabilities are: $\Pr(\click|\view)$, the probability a user clicks on a story after viewing it, and $\Pr(\vote|\click)$, the probability a user votes for a story when seeing the full content by clicking its url.

%\noteTH{I removed use of ``quality'', defined as $\Pr(\vote|\click)$, and instead use the expression $\Pr(\vote|\click)$ in the text. Is that ok? Or do we need a short term to refer to that expression in the text? If so, probably useful to call it ``quality'' (as an operational defn.), since this will help readers relate to results of the WWW paper on use of informed users.}

% no longer define quality, since not important for this paper
%We distinguish these factors as story appeal and quality. Specifically, we operationally define \emph{appeal} $r_s$ of a story $s$ is the likelihood a user who views its title and summary will vote for it: $\Pr(\vote|\view)=r_s$ (whether or not they click on its url). The \emph{quality} of $s$ is the likelihood a user votes for it after clicking its url to see the full story: $\Pr(\vote|\click)=q_s$.

Since the experiments do not record which stories users view, we use the model to estimate $\Pr(\view)$ and hence determine $\Pr(\click|\view)$ from observed url clicks. However, our experiments do record when users click on a story's url. In these cases, we know that the user viewed the story, clicked on the url, and then whether that user subsequently voted for it. This allows directly estimating \remove{quality, }$\Pr(\vote|\click)$ from the data.
These estimates for the probabilities allow determining how users respond to variations in content. % (see \sectA{quality}).
Specifically, we find that users tend to click on appealing stories:
the correlation of $\Pr(\click|\view)$  with $\Pr(\vote|\view)$ is $0.46$, which is unlikely to arise if there were no correlation ($p$-value less than $10^{-4}$ with Spearman rank test). This suggests that users evaluate an item's appeal based on its title~\cite{krumme12} (and summary when available) and then decide to proceed further with the appealing items, either by getting more information about them from the full text or immediately recommending them. This underscores the value of ``first impressions'': users generally devote less effort to items whose titles (and summaries) are less appealing. In particular, stories with the lowest quartile of appeal are, on average, less likely to get url clicks than stories with higher appeal.

The significance of basing most decisions on just a title (and summary) of a story depends on how well the title reflects the full content.
To evaluate this connection, we estimated $\Pr(\vote|\view)$ from all the url clicks in our experiments\footnote{These experiments involved 3498 users, of whom 816 clicked on at least one url.}.
Appeal has $42\%$ correlation with $\Pr(\vote|\click)$, indicating user response to story titles also somewhat estimates their response to the full content.
% from 50 and 100-story experiments, including users with >20 votes
% details of estimation of $\Pr(\vote|\view)$ are in WWW paper
% correlations between appeal, P(vote|view), and quality, P(vote|click)
%  0.19 for vetted users in 100-story no-influence experiments, P(vote|click) have large error bars
%  0.25 for all users in 50 and 100-story no-influence experiments
%  0.42 for all users with cross validation (reducing noise in P(vote|click) estimates)

\section{Results}
The social signals, which show users how many votes the stories received from prior users, are highly variable. %values (as described in \sectA{S1}).%signal values}).
To determine how user votes respond to these social signals, we must account for other factors that significantly affect votes. We quantify the effect of two such factors: story's content and its position in the list shown to a user.

\subsection{Social Signals and Individual Behavior}
Social signals could affect behavior in two ways.
First, by conveying information about the preferences of others, social signals may affect which stories users attend to. Second, they may change individual preferences in users' evaluation of stories~\cite{krumme12}.
To discriminate between these possibilities, we compare which stories users choose to evaluate and, of those, which they vote for in experiments with and without social signals.

\paragraph*{Changes in Attention}
We cannot directly measure attention. Instead, we indirectly estimated it by measuring which stories users choose to evaluate by investing the effort in reading the full text of its article. Specifically, we used url click data to examine how social signal changes $\Pr(\click|\view)$,
the probability a user clicks on a story's url to read the full text of the story after viewing its title and summary.
We used the model (\sect{model}) to adjust for position bias when determining $\Pr(\click|\view)$ in the absence of the social signal.

For each influence experiment we considered behavior starting after the 50th user, when there is a significant history of prior votes on the stories. For each user, we divided the stories into two groups: \emph{i}) those whose signal is less than the median among the social signal values shown to that user, and \emph{ii}) those with larger signals. Combining the expected and actual estimates of $\Pr(\click|\view)$ for these groups gives values grouped by the relative strength of the signal for each user.

\begin{figure}[t]
\centering \includegraphics[width=\figwidth]{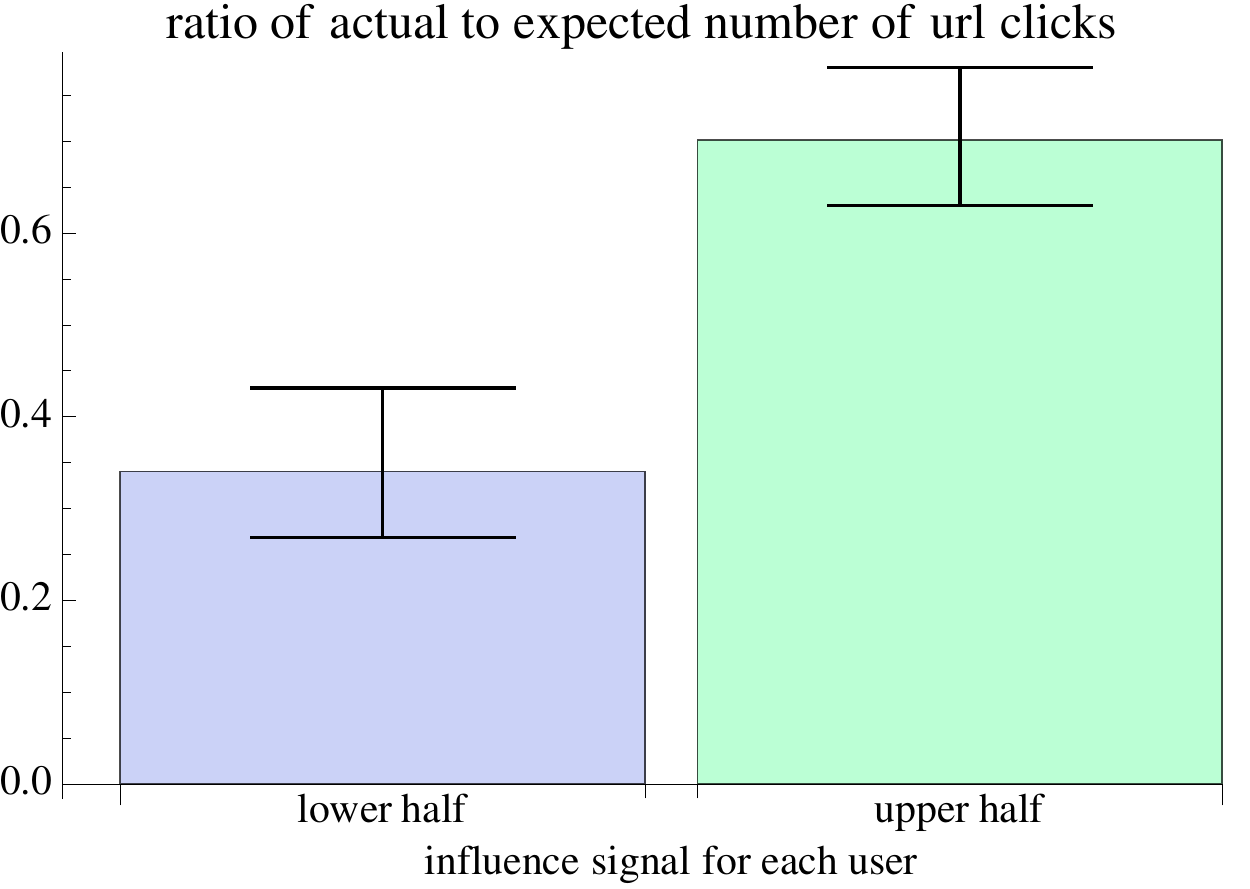}
\caption{Ratio of actual to expected url clicks for stories shown with low or high social signals to users (excluding the first 50 users in each experiment). Error bars indicate 95\% confidence intervals of the ratios. There were 28713 and 32587 instances in the lower and higher groups, respectively.
}
\figlabel{P(click|view)}
\end{figure}

\fig{P(click|view)} shows the ratio between actual url clicks and those expected when there is no social influence. The figure shows two behaviors. First, the ratios are less than one, indicating users tend to click on fewer stories with social influence compared to experiments without. This indicates a change in user effort, as discussed below. Second, the ratio is larger for larger signals, i.e., a user is relatively more likely to click on stories with larger social influence signals.
Furthermore, the correlation between $\Pr(\click|\view)$ with and without influence \remove{(both estimated as described in \sectA{P(click|view)})}
is $0.48$, which is nonzero ($p$-value less than $10^{-5}$ with Spearman rank test). This indicates that users tend to click on the same stories in the two treatments, but with considerable variation.

\paragraph*{Changes in Preferences}
We determine how social signals alter individual preferences for stories from changes in probability to recommend a story conditioned on clicking on that story's url, $\Pr(\vote|\click)$.  In the absence of social signals, users recommend half of the stories they click on, $\Pr(\vote|\click)=50\%$. This probability is higher with social signals: 66\%.
% added the following sentence to avoid seeming to miss giving statistical significance to the difference in average: pairwise test of per-story difference, given below, is a more sensitive test
%We evaluate the significance of this difference in average behavior on a per-story basis.

\begin{figure}[t]
\centering \includegraphics[width=\figwidth]{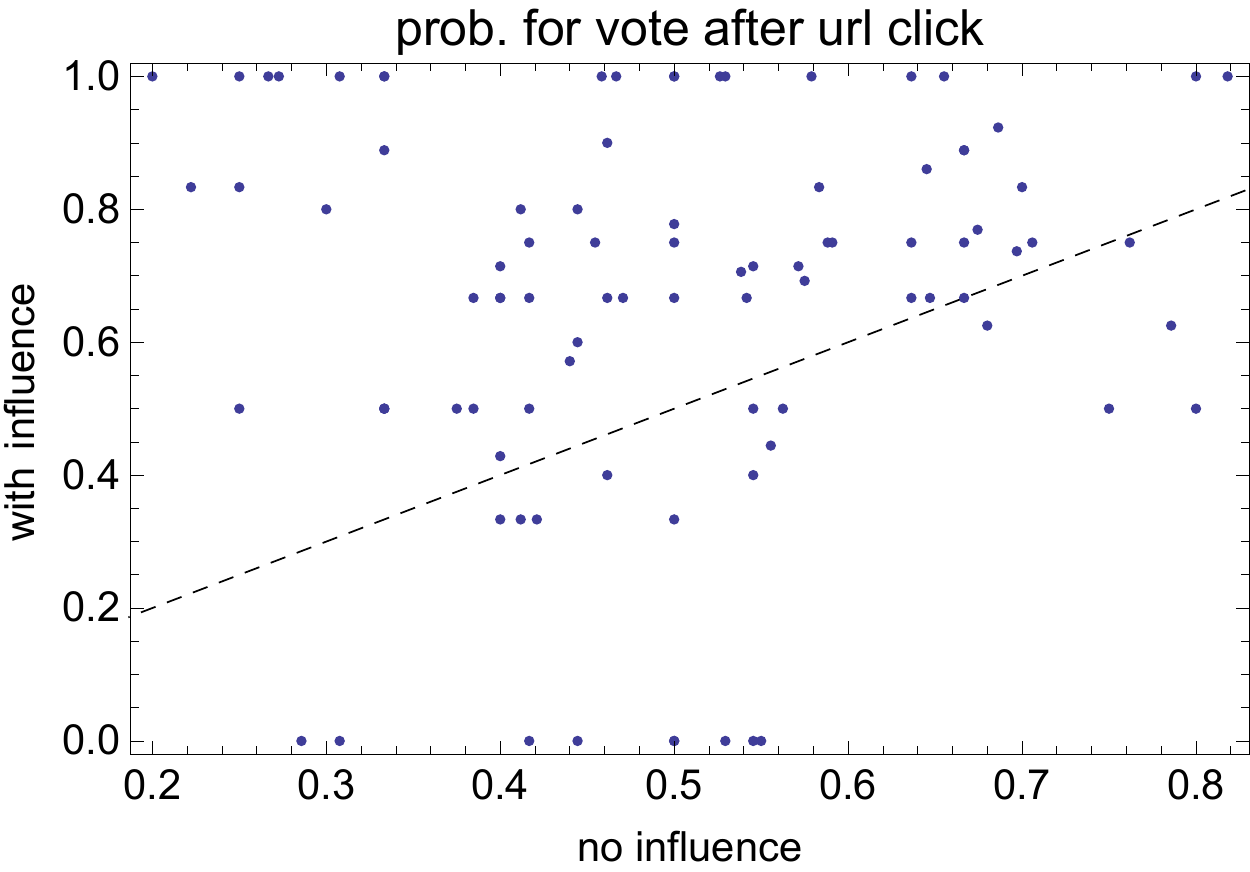}
\caption{Fraction of users who vote on each story, given that the user has clicked on the story's url, for the experiments with and without social influence. The dashed line corresponds to equal values for experiments with and without social influence.
}
\figlabel{P(vote|url)}
\end{figure}

This increase in average $\Pr(\vote|\click)$ could arise from either of two changes in user behavior. First, social influence could change preferences, increasing $\Pr(\vote|\click)$ for stories users click on and leading to the larger average value we observe for those stories. Second, as described above, users click on fewer urls when there is an social signal. If the stories they do click on tend to be more interesting than those clicked on without social influence, the \emph{average} value of $\Pr(\vote|\click)$ for stories receiving clicks will be larger even if preferences do not change, i.e., if $\Pr(\vote|\click)$ values for each story remain the same. In this case, the change in average arises from the selection of stories users click on.

We distinguished between these possibilities by examining, on a per-story basis, $\Pr(\vote|\click)$ estimated by the fraction of all url clicks for a story that produce a vote for that story. \fig{P(vote|url)} shows that most stories are more likely to receive a vote after a click when there is an social signal. Quantitatively, comparing the $\Pr(\vote|\click)$ on a per-story basis shows a significant difference (pairwise t-test $p$-value $5\times 10^{-6}$). This indicates that changing preferences rather than selection gives rise to the larger average value of $\Pr(\vote|\click)$ for all stories.

Moreover, the difference in preference increases for larger social signal values. Specifically, for stories shown with influence signals below and above the median, we have $\Pr(\vote|\click)=65\%$ and $75\%$, respectively, averaged over all stories. These differences in average values are significant on a per-story basis (pairwise t-test $p$-value is less than $10^{-4}$ for both cases).
% 10^{-4} for signals below median,  3*10^{-6} for signals above median
In other words, users are more likely to vote for a story they read when they see that it already received many votes.

\remove{ % removed as minor point to save space; also unclear how useful it is to distinguish *which* stories get votes from the change in prob. for a vote on a per-story basis discussed above
Social influence not only increases the probability of vote but also changes which stories receive votes: there is essentially no correlation between $\Pr(\vote|\click)$ for influence and no-influence cases: the correlation $0.1$ is consistent with no correlation (Spearman rank test $p$-value $0.5$).
}

%\subsection{Social Influence and Outcomes of Peer Recommendation}
\subsection{Social Signals and Collective Behavior}
We examined how social signals affect collective behavior in peer recommendation by comparing the outcomes, i.e., the number of votes stories receive, in the experiments with and without social influence.

\paragraph*{Inequality of Outcomes}
Stories differ in appeal; hence, when attention is distributed uniformly (as in the random ordering policy), we expect their popularity to vary in proportion to their appeal. Orderings that direct user attention toward the same set of stories increase the inequality of popularity~\cite{lerman14as}. Social signals can further focus user attention, increasing inequality even more.

\begin{figure}[t]
\centering \includegraphics[width=\figwidth]{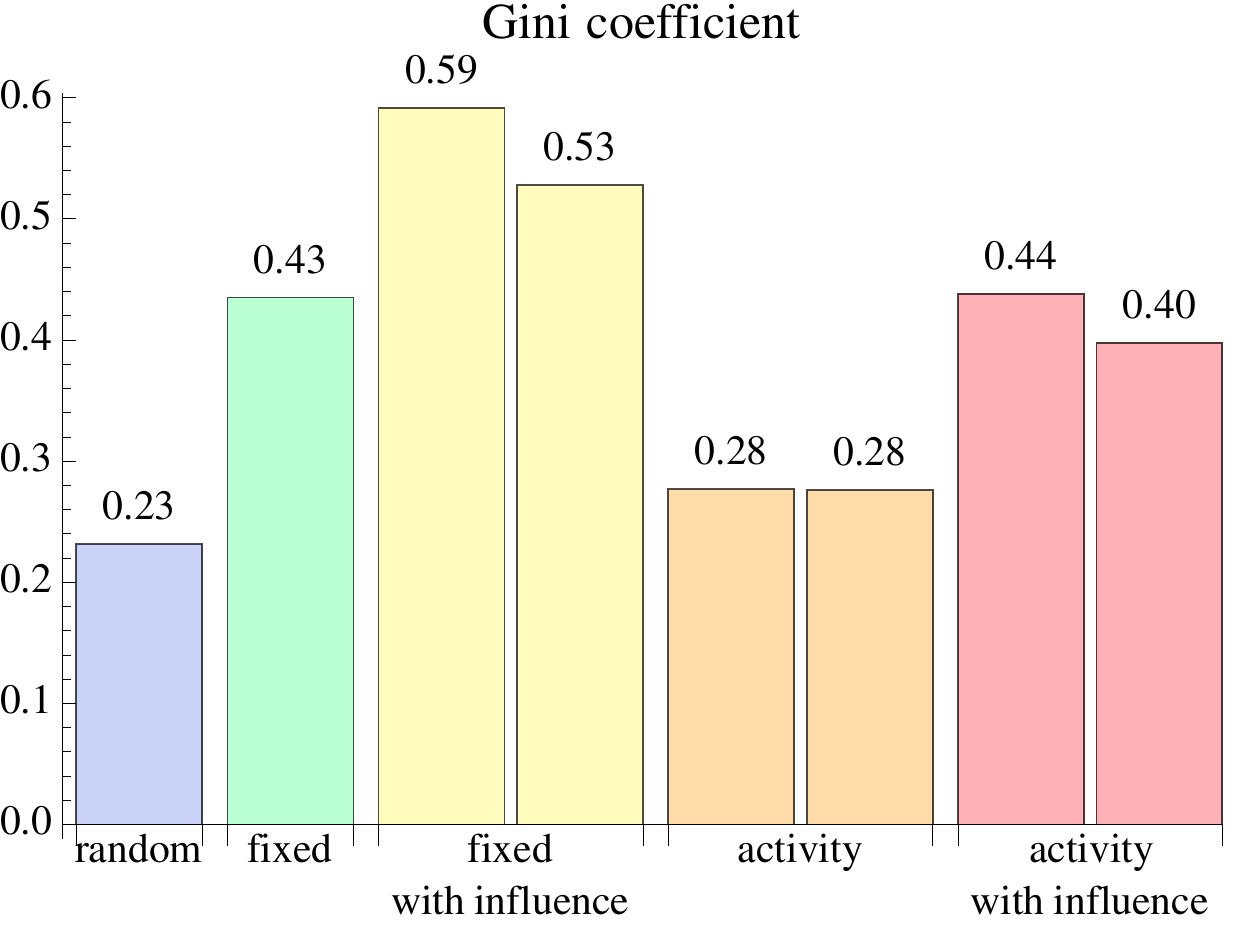}
\caption{Gini coefficient showing inequality of the total votes received by stories with different ordering policies.}
\figlabel{Gini}
\end{figure}

We use the Gini coefficient to quantify the variation in popularity:
\begin{equation}
G = \frac{1}{2S} \sum_{i,j}  \left| f_i - f_j \right|
\end{equation}
where $S=100$ is the number of stories and $f_i$ is the fraction of all votes that story $i$ receives, so $\sum_i f_i=1$.
\fig{Gini} shows the values of the Gini coefficient in our experiments. The random ordering indicates the inequality due to variation in story appeal. The remaining conditions indicate the increase in inequality due to position bias and social signal. The figure shows that social influence increases inequality, for both the fixed and activity orderings.

To quantify the significance of the increase in inequality, we focused on the experiments with fixed ordering and no social influence. In this case, each user faces the same situation so each user's behavior is independent of choices made by other users. This lack of history-dependence allows estimating the distribution of Gini coefficients from bootstrap samples (with replacement) of the users in the experiment. Each sample, with the same number of users as the actual experiment but different distributions of votes among stories, gives a corresponding Gini coefficient. All of 1000 such samples had Gini coefficient less than both of the values for the fixed ordering with social influence. Thus the Gini coefficients with influence are unlikely to arise if, in fact, there were no increase in inequality due to the social signal.

The activity ordering varies the story order based on user responses. Thus user behaviors are not independent and bootstrap samples, based on assuming independence, may not indicate the distribution of Gini coefficients that would arise without social influence. Instead, we see there is very small difference in the Gini coefficients of the two parallel worlds for the activity ordering without social influence (\fig{Gini}). By comparison, the observed values for activity with social influence are considerably larger, suggesting those values are unlikely to arise if the social signal had no effect on inequality.

\paragraph*{Reproducibility of Outcomes}
For the activity ordering, the correlations between votes in the two parallel worlds are $0.78$ and $0.63$ without and with social influence, respectively.
For the fixed ordering without social influence we considered all users part of the same ``world''. Nevertheless, due to the lack of history-dependence in this ordering, arbitrarily splitting the users in this experiment into two subsets gives, in effect, parallel worlds for this ordering. This gave correlation $0.87$ between the worlds, compared with $0.90$ for the fixed ordering with social influence.
These correlations between parallel worlds with and without social signal are not significantly different ($p$-value 0.5 for Spearman rank test).
Thus, social signal does not appear to significantly degrade the reproducibility of outcomes in peer recommendation.

\paragraph*{Strength of Social Signal and Outcomes}
To separate the effects of social signal from those of position bias and story appeal, we used the model (see \sect{model})
to estimate the expected number of votes stories would receive, based on their position and appeal, in the absence of a social signal. Comparing these estimates with the observed votes indicates how social signals change the outcomes of peer recommendation.

Quantifying the response as a function of the \emph{size} of the social signal requires identifying how users attend to social signals.
One possibility is users respond independently to each signal's value. Alternatively, users could compare signals and focus on the rank of a story's signal among all, or a sample, of the values shown. Another possibility is that users respond to signals that deviate by at least several standard deviations from the average signal value.
In our experiments, these measures of signal strength are highly correlated (over 80\%) and give similar indication of relative importance of appeal, position bias, and social influence. So our conclusions do not depend on which of these, or similar, measures most closely reflects user behavior.

\begin{figure}[t]
\centering \includegraphics[width=\figwidth]{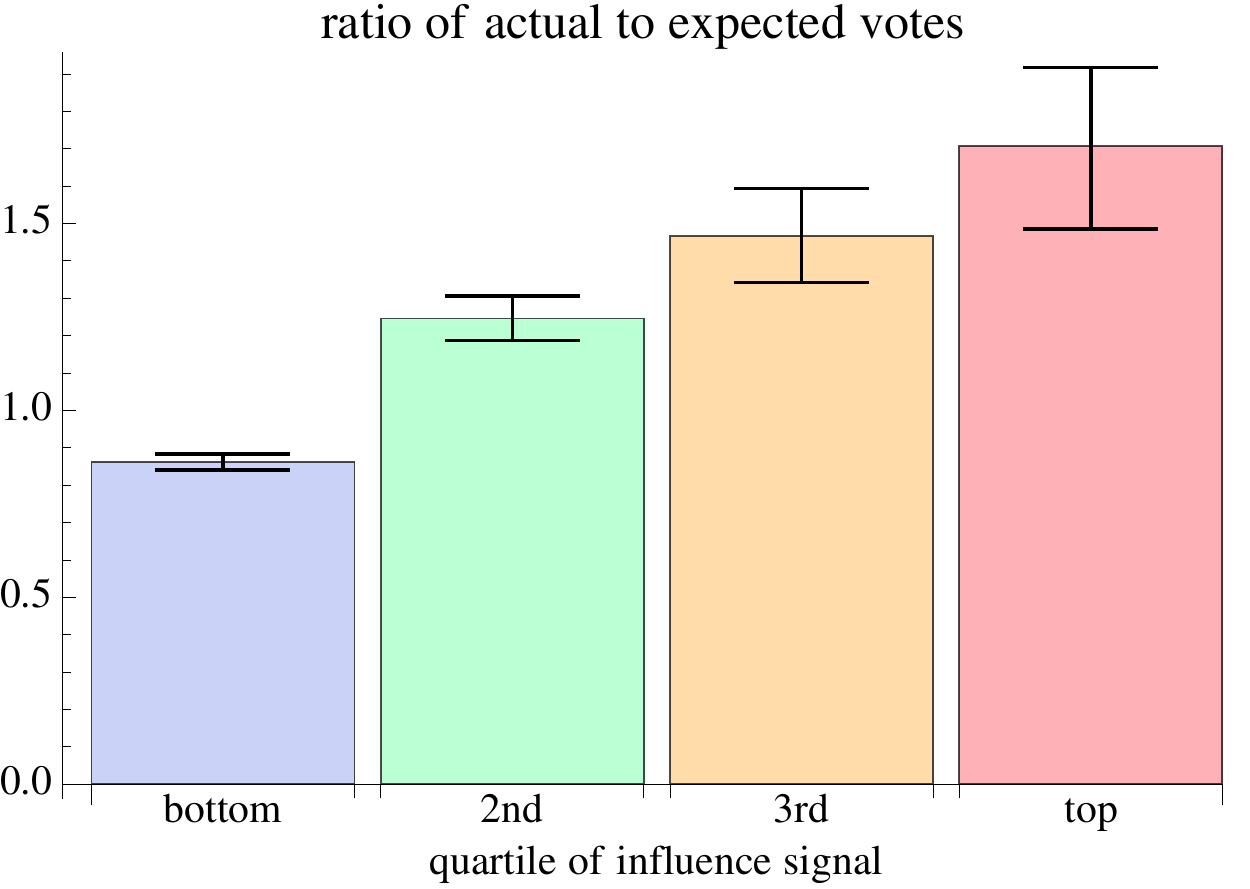}
\caption{Ratio of actual to expected votes for stories shown with each quartile of the social signal. Error bars indicate 95\% confidence intervals of the votes based on the number of instances in each quartile. There were 76110, 4273, 752 and 165 instances in the bottom to top quartiles, respectively.
}
\figlabel{response to signal}
\end{figure}

For definiteness, we focused on how votes depend on the signal value itself.
\fig{response to signal} compares observed votes to the expected numbers of votes those stories would receive when there is no social signal.
Specifically, for each story $s$ shown to a user at position $p$, we used the model to determine the probability for that user to recommend that story as $r_s v_{p}$, where $v_p$ is the probability to view a story at position $p$.
We combined these values for all stories within each quartile of the full signal range in all social influence experiments. For each quartile, this gave the actual number of votes, the expected number and the number of instances (i.e., number of times a story was shown to users with a signal in each quartile). The figure shows the ratio of actual to expected votes, along with 95\% confidence intervals estimated by treating each vote as an independent sample.
Stories associated with signals in the bottom quartile get fewer votes than expected. Those with signals in the top quartile get about twice as many votes as those in the bottom quartile.
This variation compares with the ratio of mean values in top and bottom quartiles of appeal ($r_s$) and position bias ($v_p$) of 3 and 4, respectively. Thus, social influence is responsible for about half the variation of popularity as that created by the differences in story content and position.

The increase in votes with the size of the social signal raises the question of how the effect changes with time. As more users view the stories, their votes increase the magnitude of the social signal, i.e., the number of prior votes for the stories. Thus the signals may exert a stronger effect over time. However, comparing responses by early (first 100) and late (subsequent) users in each experiment (see \sectA{response evolution}),
indicates that a larger range of signals does not lead to a significantly larger variation in response.

\paragraph*{Collective Efficiency}
Social signals reduce the effort users devote to the recommendation task. One measure of effort is a user's session time, excluding the time required to read instructions and do the post-survey~\cite{lerman14as}. % Fig.~8 of PLoS paper compares vetted vs. nonvetted users, without influence
With social signals, users spend, on average, about 40 seconds or 20\% less time on the task than users without the social signals. This difference is significant ($p$-value less than $10^{-10}$ with Mann-Whitney test).

Another measure of user effort is how often they click on a story's url.
For the roughly $25\%$ of users who click on at least one url, we find a significant difference in the number of stories they click on. Specifically, without influence, such users click on 4.3 url's, on average. With influence, they click on just 2.5, a significant reduction ($p$-value less than $10^{-4}$ with Mann-Whitney test).
%\noteKL{Mention here that users tend to click on higher appeal stories, thereby focusing their evaluation effort on higher quality stories.}

% the correlation between number of votes and appeal\remove{, i.e., $\Pr(\vote|\view)$, determined by the model} is $0.77$ and $0.79$
% Good outcomes, with expending less effort
% Focusing effort on evaluating higher appeal stories.

%\paragraph*{Story Appeal and Outcomes}
A possible consequence of reduction in effort is degraded performance of peer recommendation. To evaluate this possibility, we measure performance by how well the outcomes of peer recommendation reflect the preferences of the user community. In our case, we define preferences for the stories by their appeal.
Position bias significantly affects how well users identify appealing stories and magnifies the inequality of outcomes beyond that expected from variations in story appeal~\cite{lerman14as}. \fig{Gini} shows that social influence signals further increase inequality. This suggests that social signals, like position bias, could degrade the relation between outcomes and appeal. However, we find this is not the case.
Specifically, the correlation between number of votes and appeal\remove{, i.e., $\Pr(\vote|\view)$, determined by the model} is $0.77$ and $0.79$ for activity ordering, without and with social signals, respectively. For the fixed ordering, these correlations are $0.45$ and $0.34$. Both cases are consistent with social influence having no effect on this correlation ($p$-value $0.7$ for Spearman rank test).
% p-values 0.657 for activity, 0.668 for fixed, both round to 0.7
Therefore, social influence increases the efficiency of peer recommendation by attaining similar levels of performance as without influence, but with less user effort.

%\noteTH{Possibly add correlation between number of votes and quality, as another evaluation of whether influence reduces performance, by reducing quality of recommended stories [noted by review 2 of WSDM paper]. Though we're deemphasizing quality in this paper, so don't include??}

In summary, social signals have three effects:
\emph{i}) directing attention toward stories prior users voted for, \emph{ii}) increasing user preference for the stories whose full content users examine, and \emph{iii}) increasing the efficiency of peer recommendation by focusing user evaluation effort on higher-appeal stories.

\section{Conclusion}
Our experiments quantified how social signals affect behavior and outcomes of crowdsourcing tasks, specifically, in peer recommendation.
Similar to other studies, we observe the ``irrational herding effect''~\cite{Lorenz11,Muchnik13}, wherein a large social signal indicating popularity causes the participants to partially rely on the signal, rather than personal judgement, to determine whether the content is interesting.

The stronger the social signals, the more likely the story receives more votes than expected from its appeal and position in the user interface. However, social signal was less important than story content or position: compared to the variation in attention stories received due to their position, social signals accounted for half as much variance. While these differences did not significantly change the reproducibility of outcomes, as compared to the no influence condition, they did produce more unequal outcomes.
In addition, social signals have benefits that were not previously recognized. By reducing the evaluation effort, social signals increase the efficiency of peer recommendation.

Social signals changed not only which stories received attention, but also how they were evaluated.  We found that people tended to vote for stories that many others already found interesting.
The larger the ``peer pressure'', i.e., the larger the value of the social signal, the more likely users were to vote for the story after seeing its full content (with a url click). In addition, users devoted less effort (both in time and number of url clicks) to the task when provided with social signals. This suggests users rely, to some extent, on the social signal, rather than personal judgement, to determine whether the content is interesting. This herding effect~\cite{Lorenz11,Muchnik13} increased the inequality of number of votes among stories, well beyond that expected from variation in the appeal of the content itself and that due to position bias.

In spite of larger inequality, social influence did not significantly affect performance of peer recommendation as measured by correlation between outcomes and underlying story appeal. Moreover, this was achieved with substantially less user effort, as measured by the number of url clicks and votes. Thus, it appears that by focusing user attention on more appealing stories, social influence helps make the system as a whole more efficient.

Our experiments contrast with other examples of social influence. For instance, Netflix, YouTube or Amazon users make implicit recommendations by selecting items for their own use, rather than identifying items they think \emph{others} may find interesting. The web sites then recommend items based on what similar people liked~\cite{Koren09}. In this setting, which was experimentally investigated by the MusicLab study~\cite{Salganik06,Salganik08}, social influence plays a largely informational role~\cite{krumme12}.
Another example is content whose value to a person depends on its adoption by many others, in which case a signal of prior adoption directly affects a person's evaluation of the desirability of the content. An example of such ``bandwagon effect'' is watching a popular TV show in order to discuss it with others the next day.
An additional issue is the different types of social influence discussed in the introduction. In particular, social influence can depend on the user's relationship with the person providing the signal, e.g., a friend instead of the general user community.

The experimental design using MTurk can be extended to address additional questions on how people respond to signals of prior users' preferences. For instance, experiments could identify which aspect of social signals users primarily attend to (e.g., absolute value, rank or variation from average value for the stories), by manipulating the value of the signal shown to users. That is, unlike the experiments reported here, with this manipulation the number of votes shown for a story would not necessarily be the actual number of votes it received.
% question by reviewer 2 of WSDM paper re. experiment choice
This experimental approach could examine behavior with different types of stories. These could include topics with more subjective or variable opinions than the science stories we considered. The stories could also include a mix of professional and amateur authors. In this way, the experiment could vary the type of content to match different types of web sites.
Results of such experiments could help develop a model of recommendation incorporating social signal (cf.~\cite{Stoddard15,krumme12}),
%Greg Stoddard's paper has regression model including influence signal
and thereby suggest how peer recommendation performance would react to various choices for which signals to show users, and when.
The results of these experiments may help the development of algorithms to control collective performance~\cite{Abeliuk15}.

% Future directions
% Optimizing efficiency of collective
% Exploit heterogeneity (some people are willing to put in extra effort, but most are not) to make the collective more efficient
% Generalization - look at wider variety of topics

\subsection*{Acknowledgments}
% acknowledge funding; an unnumbered section
This work was supported in part by Army Research Office under contract W911NF-15-1-0142, by the Air Force Office for Scientific Research under contract FA9550-10-1-0569, and by the Defense Advanced Research Projects Agency under contract W911NF-12-1-0034.

%\bibliographystyle{hcj}
%\bibliography{references}
%%% -*-BibTeX-*-
%%% Do NOT edit. File created by BibTeX with style
%%% ACM-Reference-Format-Journals [18-Jan-2012].

\appendix
%Appendix A

% alternate numbering in appendix
%\setcounter{figure}{0}
%\makeatletter
%\renewcommand{\thefigure}{A\@arabic\c@figure}
%\makeatother
%
%\setcounter{table}{0}
%\makeatletter
%\renewcommand{\thetable}{A\@arabic\c@table}
%\makeatother
%
%\setcounter{section}{0}
%\makeatletter
%\renewcommand{\thesection}{A\@arabic\c@section}
%\makeatother

\section{User Recruitment and Vetting}\sectlabel{vetting}

% description of experiment uses 'recommend' not 'vote', since 'recommend' was actual word shown to users
We published experiments as tasks on Amazon Mechanical Turk to recruit participants for the study from a large pool of workers. Workers who accepted the task were instructed as follows: ``We are conducting a study of the role of social media in promoting science. Please click `Start' button and recommend articles from the list below that you think report important scientific topics. When you finish, you will be asked a few questions about the articles you recommended. (Please remember, once you finish the job, the system won't allow you to do it again).'' Workers were paid \$0.12 for completing the task and each person was allowed to participate in the experiment only once. The pay rate was comparable to similar tasks in other research studies~\cite{Kittur13,mason12} and was set low to discourage gaming Mturk.
University of Southern California's Institutional Review Board (IRB) reviewed the experiment design and designated it as ``non-human subjects research.''
%\noteTH{Instead of ``classified'', can we say the IRB ``approved'' the experiment?}
%\noteTH{could note what ``non-human subjects research.'' means in this context -- presumably a category not requiring more formal approval rather than its plain English meaning of not involving humans, which is not correct.}

Participants were shown a list of one hundred science stories, drawn from the Science section of the New York Times  and science-related press releases from major universities (sciencenewsdaily.com), which included titles and summaries.
%\noteTH{UI figure shows both title and summary for each story. Where did the summaries come from?}
The list was sufficiently long to require participants to scroll to see all stories.  They could choose to recommend a story based on the short description or click on the link to view the full story. We recorded all actions, including recommendations and URL clicks, and the position of all stories each participant saw. After a vote, the button changed color to indicate the user voted for that story. Participants were not allowed to undo their votes: subsequent clicks of the button brought up a message box reminding them to vote for a story only once. Although participants were not told ahead of time how many stories to vote for, if they tried to finish the task before making five actions (either votes or URL clicks), a message box prompted them to recommend five stories. % 'recommend' instead of the term 'vote' used throughout the paper since the message box describes the UI, which used 'recommend'

Upon finishing the task, participants were asked to name two important themes in the stories they voted for and solve a simple arithmetic question. Only those who correctly answered the arithmetic question were considered to have completed the task and paid.
Similar to earlier experiments~\cite{lerman14as}, we used a multi-step strategy to reduce spam and to weed out unmotivated workers. First, we selected workers using qualifications provided by Mturk: they lived in the US, had completed at least 500 tasks on Mturk, and had a 90\% or above approval rate. In addition, we ignored actions by participants who recommended more than 20 stories. Their votes did not affect the social signals or the order of stories.

\tbl{summary} summarizes the experiments, showing the number of participants (i.e., users), the number of votes, and url clicks for each condition. The random ordering policy is the control condition used to identify story appeal and the effect of position bias.

\begin{table}[ht]
% from "visibility experiments summary-2"
\centering
\caption{Summary of experiments. The history-dependent ordering (activity) and the social influence treatments each have two independent experiments.
}\tbllabel{summary}
\begin{tabular}{lccc}
ordering\remove{ policy} 	& users 	& votes 	& url clicks \\ \hline
random	& 199		& 1873 	& 164 \\ 	% combine two experiments: 70+129 users, 729+1144 votes, 67+97 url clicks
fixed		& 217		& 1978	& 424 \\ % combine two experiments: 64+153 users, 634+1344 votes, 84+340 url clicks
activity	& 286 \& 193		& 2586 \& 1764		& 246 \& 247 \\
\hline \multicolumn{4}{c}{with social signal} \\
fixed		& 192 \& 211	& 1572 \& 2057	& 125 \& 135 \\
activity	& 200 \& 210	& 1892 \& 1959	& 113 \& 144 \\
\hline
{\sc total}	& 1708	& 15681	& 1598 \\
\end{tabular}
\end{table}

\section{Response to Social Signals}
\sectlabel{response evolution}

\begin{figure}[t]
\centering \includegraphics[width=\figwidth]{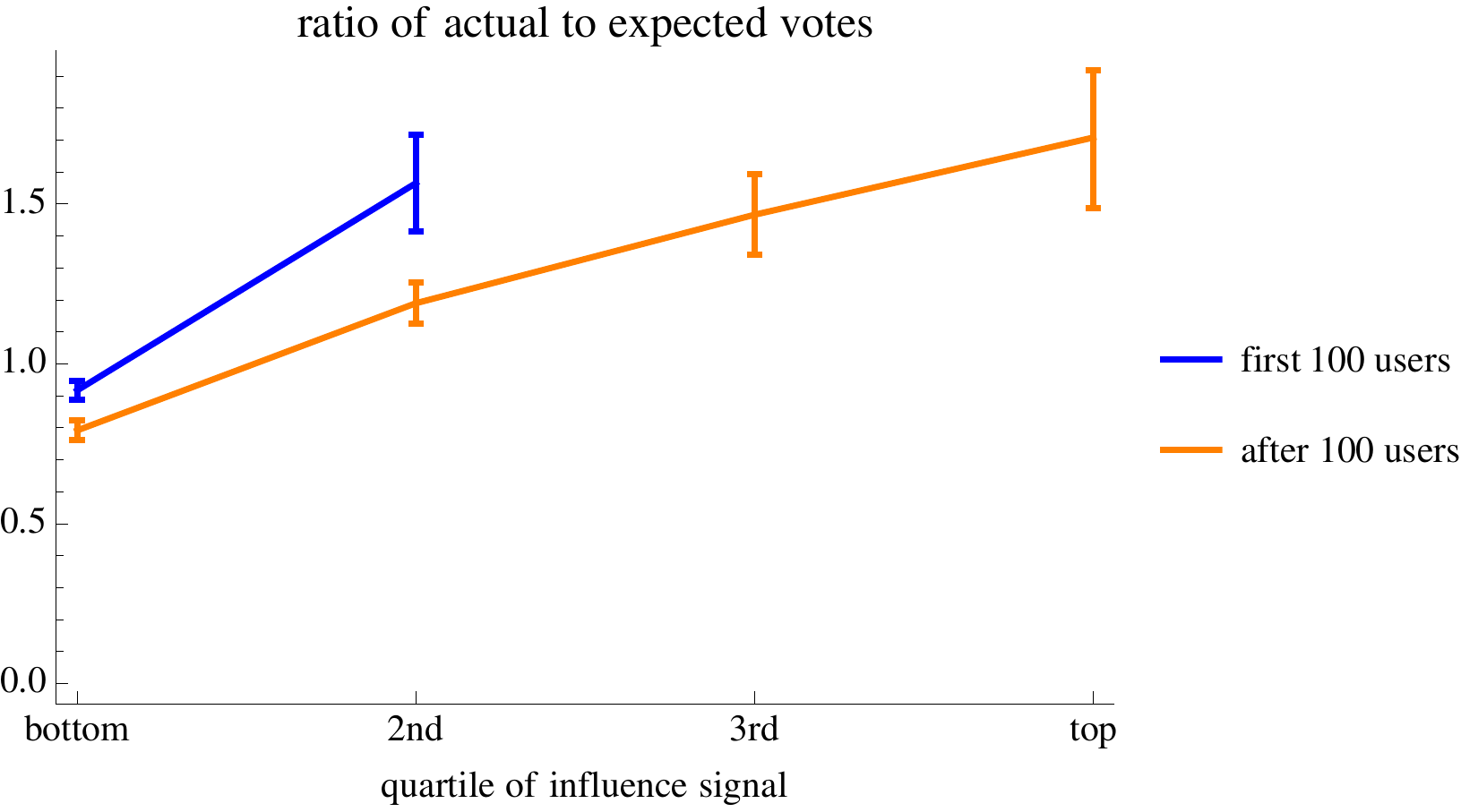}
\caption{Ratio of actual to expected votes for stories shown with each quartile of the social influence signal to the first 100 users and the rest of the users in each experiment. Error bars indicate 95\% confidence intervals of the votes based on the number of instances in each quartile. By comparison, \fig{response to signal} shows the same values but aggregated over all users.}
\figlabel{response by early vs late users}
\end{figure}

As experiments progress, the social signal values get larger and distributed more broadly. % (\fig{signal distribution}).
If users respond primarily to the magnitude of the signal, the effect of social signal would tend to become larger relative to other factors, e.g., position bias, as stories accumulate votes. To test this possibility, we compare response to the signal for the first 100 users in each experiment with those of the rest of the users. Specifically, \fig{response by early vs late users} shows the response measured by the ratio of actual to expected votes for each quartile of signal value occurring in our experiments. Early users encounter relatively small signals, in all cases within the bottom two quartiles of signal values appearing in the experiments. Nevertheless, such users have a similar change in response between the highest and lowest signals they encounter as shown by subsequent users who encounter the full range of signal values.

% "Model Evaluations" section could be supplementary material: unlike the model description,  the model evaluation is not directly used in the paper
\section{Model Evaluations}
\sectlabel{evaluations}
To evaluate the model described in \sect{model}, we use it to predict the number of votes stories receive under different ordering policies in the no-influence condition experiments.
Specifically, we use  all no-influence experiments other than those with the random ordering policy (which were used to estimate model parameters). This \emph{model test data} contains votes by 1319 users~\cite{lerman14as}.
% 1518 total users for non-influence experiments, of which 199 with random interface used to estimate model parameters, so the remaining 1319 for testing

In addition to the fixed and activity orderings used in the social influence experiments, this test data includes the remaining orderings from the prior experiments~\cite{lerman14as}.
The \emph{popularity} ordering presented stories in decreasing order of the number of recommendations they had received. Popularity-based ordering is widely used by web sites to highlight interesting content.
This ordering produced highly variable and unpredictable outcomes~\cite{lerman14as}, because it tended to focus user attention on the same set of highly recommended stories, which became even more popular.
The \emph{reverse} policy inverted the order used with the fixed policy.
% mention parallel worlds since figure comparing predicted and actual votes shows activity and popularity twice; without this note, that duplication is not explained without reading PLOS paper
For each of activity and popularity orderings, we conducted two `parallel world' experiments.

%\subsection{Baselines}

We compare predictions of the model with three alternative baselines: \model{appeal}, \model{position bias} and \model{random}.
In the \model{appeal} model, position bias plays no role so users recommend stories based solely on how appealing they are. That is, the probability to recommend a story is independent of the position where that story is shown to a user, so \eq{prob(recommend)} becomes $\rho(s,p)=V r_s$, where $V$ is a constant, equal to $ \left<v_p\right>$, the average value of $v_p$ over all positions.

In the \model{position bias} model, users recommend stories based solely on their position, so $\rho(s,p)=R v_p$ where $R$ is a constant, equal to $ \left<r_s\right>$, the average value of $r_s$ over all stories.

In the \model{random} model, each story is equally likely to be recommended: $\rho(s,p)$ is a constant value, independent of story $s$ and the position $p$ that story is shown to a user.

These alternative models allow us to evaluate the relative importance of story appeal and position bias in producing the behavior observed in the experiments.

\paragraph{Aggregate Response}
%\tbl{overall prediction} examines how
One measure of the model is how well it predicts the aggregate response, i.e., the total number of stories recommended by users under different ordering policies.
We use relative error, the difference between actual and predicted recommendations, divided by the predicted recommendations, to measure prediction accuracy.
%The relative errors for the ordering policies in different parallel worlds experiments were: $-7\%$ \& $-2\%$ (fixed), $-7\%$ \& $-5\%$ (activity), and $-13\%$ \& $-11\%$ (popularity).
The model overestimates the total response by about $10\%$, with somewhat larger error for the popularity ordering than for the others.

\begin{table}[ht]
\centering
%\caption{Prediction accuracy for total number of votes (popularity): the relative error is the difference between actual and predicted recommendations, divided by the predicted recommendations.}
\caption{Relative error of the predicted number of votes\remove{ (popularity)}.}
\tbllabel{overall prediction}
\begin{tabular}{lc}
ordering\remove{ policy} 	&  relative error 	 \\ \hline
% list each interface individually, to show experiments involve testing various interfaces, including with history dependence.
fixed		& $-7\%$	\\ % combine two experiments: 64+153 users, 634+1344 votes
reverse	& $-2\%$	\\
activity	& $-7\%$ \& $-5\%$		 \\
popularity	& $-13\%$ \& $-11\%$	\\
\end{tabular}
\end{table}

\paragraph{Response to Stories}

\begin{figure}[!ht]
\centering
\includegraphics[width=\figwidth]{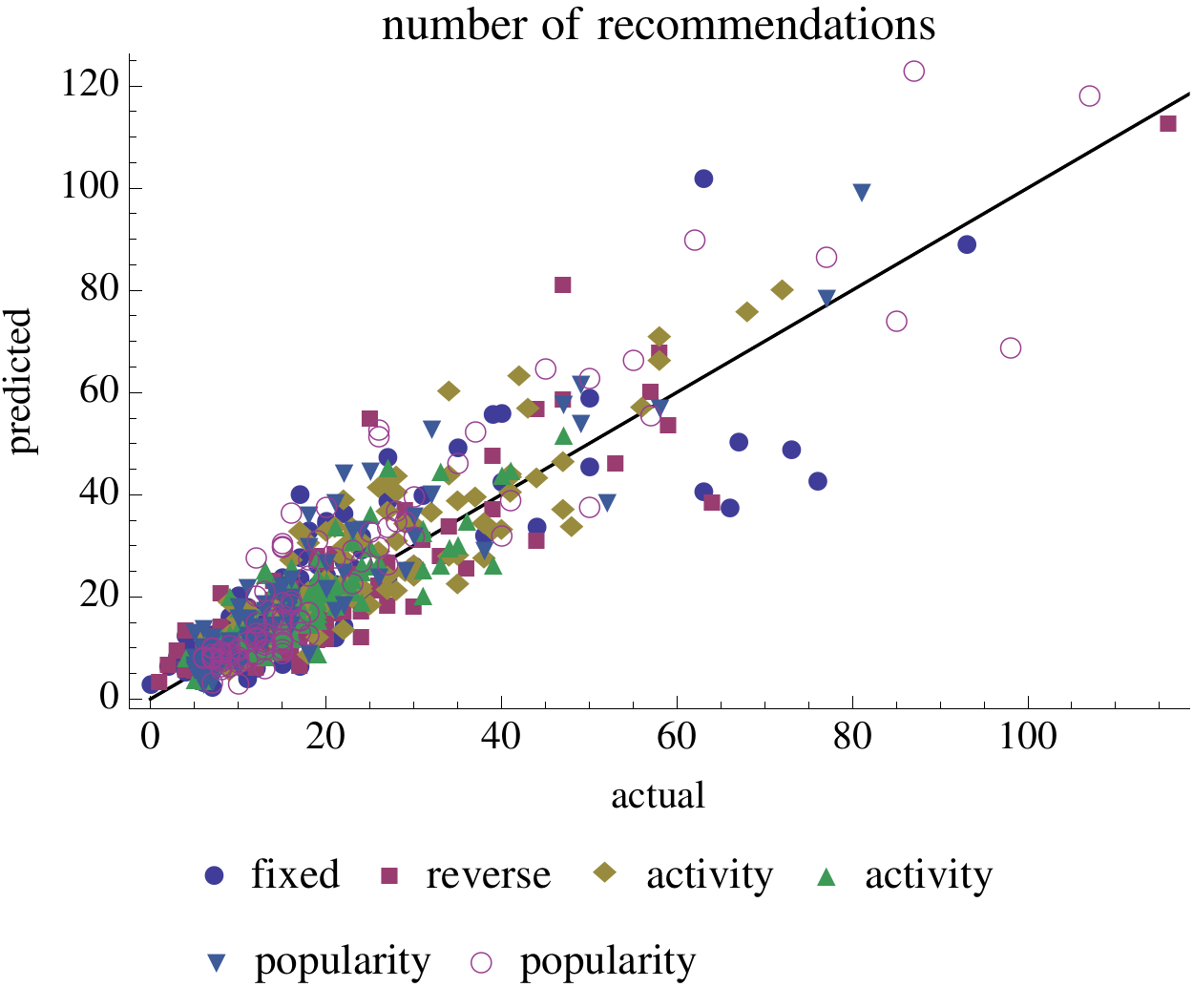}
\caption{Prediction vs. actual number of votes for the stories for each ordering policy. The line indicates where the predicted and actual numbers are the same.}
\figlabel{prediction}
\end{figure}

\fig{prediction} compares predicted and actual number of votes\remove{, i.e., story popularity,} on all stories under different ordering policies.
When different users see story $s$ at positions $p_1,p_2,\ldots$, the expected number of votes, i.e., the model's prediction, is $E(s)=\sum_k \rho(s,p_k)$ with $\rho(s,p)$ given by \eq{prob(recommend)}.

\tbl{prediction} quantifies the prediction accuracy for each ordering policy.
This shows the model predicts the rank ordering of the number of recommendations fairly well. This could be useful for producing a ranked list of items, e.g., ``best sellers'' or ``top hits'', rather than predicting their exact popularity.
In addition, \tbl{prediction} shows the parallel worlds for the history-dependent orderings have consistent prediction errors.

\begin{table}[ht]
\centering
\caption{Prediction accuracy of the full model. The second column gives the rank correlation between actual and predicted votes. The third column is the mean value of the relative error, i.e., absolute value of difference between actual and predicted votes, divided by the prediction. The last column is the fraction of stories whose votes are within two standard deviations of the predicted value. For comparison, if votes were independent, the central limit theorem gives $95\%$ for this fraction, which is close to the observed fractions.
}\tbllabel{prediction}
{
\begin{tabular}{@{}lccc@{}}
%	& \multicolumn{2}{c}{number of}\\
ordering\remove{ policy} 	& rank correlation 	& relative error 		& fraction \\ \hline
fixed		& $0.85$			& $0.40$			& $0.90$\\ % combine two experiments: 64+153 users, 634+1344 votes
reverse	& $0.81$			& $0.31$			& $0.95$\\
activity	& $0.85$ \& $0.86$	& $0.25$ \& $0.22$	& $0.98$ \& $0.98$ \\
popularity	& $0.89$ \& $0.90$	& $0.27$ \& $0.29$ 	& $0.95$ \& $0.91$\\
\end{tabular}
}
\end{table}

In terms of the quantitative accuracy, predictions are usually within about $30\%$ of the observed values, with some bias toward predictions above the actual values. The errors are similar for all the interfaces, indicating the model can compare outcomes irrespective of presentation order.

% measure of prediction error compared to variance predicted by the model
Even if the model's predictions were accurate on average, there would be statistical errors. A quantitative measure is the number of standard deviations between the actual and expected values, i.e., $|A(s)-E(s)|/\sqrt{V(s)}$ for story $s$, where $V(s)$ is the variance in number of recommendations predicted by the model.
The variance $V(s)$ arises from variations in votes given the values of $\rho(s,p_s)$, as well as from variation in the $r_s$ and $v_p$ values, indicated by the error bars in \fig{interestingness}. Variations in $r_s$ and $v_p$ values are correlated, precluding evaluating $V(s)$ by assuming independent variations. Instead we created 1000 samples from an approximation to the joint distribution of these values, with each sample a set of $r_s$, $v_p$ values for all stories and positions, respectively. Specifically, the full distribution is proportional to the likelihood of the training set $\exp(\Lfull)$\remove{, including the regularization}. Our approximation is the multivariate normal distribution matching the quadratic expansion of $\Lfull$ around its maximum. This captures the correlation among the values and closely matches the full distribution around its maximum, i.e., for values contributing to most of the probability.
We simulated a set of votes on the stories for each sample. For story $s$, the average number of votes in these samples is close to $E(s)$ and their variance gives our estimate of $V(s)$\remove{, accounting for the correlation in the parameters}.
\tbl{prediction} shows that over $90\%$ of stories are within two standard deviations of the prediction. Thus the model not only predicts the outcomes, but the variance from the model indicates the likely accuracy of the predictions.

For comparison, \tbl{prediction alternatives} shows the relative errors for the three baseline models: \model{appeal}, \model{position bias} and \model{random}. The two baselines that ignore position bias give larger errors, whereas the model accounting only for position bias gives similar error. This indicates position bias is a major contributor to the variation in number of recommendations stories receive.

\begin{table}[t]
\centering
\caption{Mean value of relative error in popularity predicted using baseline models.
}\tbllabel{prediction alternatives}
{
\begin{tabular}{@{}lccc@{}}
	& \multicolumn{3}{c}{model} \\
ordering 	& \model{appeal}	 & \model{position bias} 	& \model{random} \\ \hline
fixed		& $0.59$			& $0.36$			& $0.61$\\
reverse	& $0.55$			& $0.40$			& $0.55$\\
activity	& $0.30$ \& $0.27$	& $0.27$ \& $0.28$	& $0.39$ \& $0.39$ \\
popularity	& $0.47$ \& $0.45$	& $0.24$ \& $0.23$ 	& $0.61$ \& $0.62$\\
\end{tabular}
}
\end{table}

\paragraph{Which Stories Users Vote For}
Our model assumes a homogeneous user population and that users consider stories independently. As discussed above, this simplification allows predicting how many votes a story receives, but prevents the model from predicting the number of votes a specific user will make.
%\noteTH{to save space, could remove next sentence as unnecessary detail}
In particular, user choices differ significantly from independence since we prompt users to make at least 5 votes if they make too few and we restrict consideration to users with at most 20 votes~\cite{lerman14as}. % alternative to this citation is refer to section on vetting

Nevertheless, the model does indicate \emph{which} stories a user votes for when he or she makes a specified number of votes, $m$, for a specific story ordering. Specifically, \eq{prob(recommend)} gives the probability for voting for each story.
%
%We compare our model with the baselines \model{appeal}, \model{position bias} and \model{random}.
A simple measure of prediction quality is the fraction $f$ of the user's $m$ actual votes that are among $m$ stories with the largest predicted vote probabilities. In the \model{random} model, each of the ${n \choose m}$ ways to pick $m$ stories from among the $n=100$ stories in our experiments is equally likely. The probability that exactly $k$ of these choices match the user's actual $m$ votes is
\begin{equation}
{m \choose k} {n-m \choose m-k} / {n \choose m}
\end{equation}
Thus the expected value of the fraction $f$ in the \model{random} model is $m/n$.

\tbl{user prediction} shows the median value of $f$ for the model test set.
Thus, accounting for story appeal to the user community identifies considerably more of users' votes than random-selection, and additionally accounting for position of stories shown to each user improves this prediction.

\begin{table}[t]
\centering
\caption{Median fraction of user votes among the top predicted stories for each model.}\tbllabel{user prediction}
\begin{tabular}{lccc}
				&		& \multicolumn{2}{c}{95\% confidence}\\
model			& median	& \multicolumn{2}{c}{interval} \\ \hline
\model{full}		& 0.27	& 0.25	& 0.28 \\
\model{appeal}	& 0.18	& 0.17	& 0.20 \\
\model{position bias}	& 0.27	& 0.25	& 0.29 \\
\model{random}	& 0.07	& 0.07	& 0.08 \\
\end{tabular}
\end{table}

\end{document}